\documentclass[runningheads,a4paper]{llncs}

\usepackage{amssymb}
\usepackage{amsmath}
\usepackage{array}
\usepackage{multirow}
\usepackage{rotating}
\usepackage{graphicx}
\usepackage{comment}
\usepackage[square,numbers,comma]{natbib}

\usepackage{url}
\urldef{\mailsa}\path|{alfred.hofmann, ursula.barth, ingrid.haas, frank.holzwarth,|
\urldef{\mailsb}\path|anna.kramer, leonie.kunz, christine.reiss, nicole.sator,|
\urldef{\mailsc}\path|erika.siebert-cole, peter.strasser, lncs}@springer.com|    
\newcommand{\keywords}[1]{\par\addvspace\baselineskip
\noindent\keywordname\enspace\ignorespaces#1}

\begin{document}

\mainmatter  

\title{High-Performance Pseudo-Random Number Generation on Graphics Processing Units}

\titlerunning{High-Performance PRNG on GPUs}

%
%
\author{Nimalan Nandapalan\inst{1} %
    \and Richard P. Brent\inst{1,2} %
    \and Lawrence M. Murray\inst{3} %
    \and\\ Alistair Rendell\inst{1}}
\authorrunning{Nandapalan et al.}

\institute{Research School of Computer Science, The Australian National
  University 
\and Mathematical Sciences Institute, The Australian National University
\and CSIRO Mathematics, Informatics and Statistics}

%
%

\toctitle{Lecture Notes in Computer Science}
\tocauthor{Authors' Instructions}
\maketitle

\begin{abstract}
This work considers the deployment of pseudo-random number generators (PRNGs) on
graphics processing units (GPUs), developing an approach based on the
xorgens generator to rapidly produce pseudo-random numbers 
of high statistical
quality. The chosen algorithm has configurable state
size and period, making it ideal for tuning to the GPU architecture. We
present a comparison of both speed and statistical quality with
other common parallel, GPU-based PRNGs, demonstrating favourable
performance of the xorgens-based approach.
\keywords{Pseudo-random number generation, graphics processing units, Monte Carlo}
\end{abstract}

\section{Introduction}

Motivated by compute-intense Monte Carlo methods, this work considers the
tailoring of pseudo-random number generation (PRNG) algorithms to graphics
processing units (GPUs). Monte Carlo methods of interest include Markov chain
Monte Carlo (MCMC)~\cite{Gilks1995}, sequential Monte Carlo~\cite{Doucet2001}
and most recently, particle MCMC~\cite{Andrieu2010}, with numerous
applications across the physical, biological and environmental sciences. These
methods demand large numbers of random variates of high statistical
quality. We have observed in our own work that, after acceleration of other
components of a Monte Carlo program on GPU~\cite{Murray2011,Murray2011a}, the
PRNG component, still executing on the CPU, can bottleneck the whole
procedure, failing to produce numbers as fast as the GPU can consume them. The
aim, then, is to also accelerate the PRNG component on the GPU, without
compromising the statistical quality of the random number sequence, as
demanded by the target Monte Carlo applications.

Performance of a PRNG involves both speed and quality. A metric for the former
is the number of random numbers produced per second (RN/s). Measurement of the
latter is more difficult.  Intuitively, for a given sequence of numbers, an
inability to discriminate their source from a truly random source is
indicative of high quality. Assessment may be made by a battery of tests which
attempt to identify flaws in the sequence that are not expected in a truly
random sequence. These might include, for example, tests of autocorrelation
and linear dependence. Commonly used packages for performing such tests are
the DIEHARD~\cite{Marsaglia1996} and TestU01~\cite{LEcuyer2007} suites.

The trade-off between speed and quality can take many forms. 
Critical parameters are the \emph{period} of the generator (the length of the
sequence before repeating) and its \emph{state size} (the amount of working
memory required). Typically, a generator with a larger state size will have a
larger period. In a GPU computing context, where the available memory per
processor is small, the state size may be critical. Also, a
conventional PRNG produces a single sequence of numbers; an added challenge
in the GPU context is to produce many uncorrelated
streams of numbers concurrently.

Existing work includes the recent release of NVIDIA's CURAND~\cite{CURAND}
library, simple random generators in the Thrust C++ library~\cite{Thrust}, and
early work for graphics applications~\cite{Howes2007}. Much of this work uses
simple generators with small state sizes and commensurately short periods, in
order not to exceed the limited resources that a GPU provides to individual
threads. The statistical quality of numbers produced by these algorithms is
not necessarily adequate for Monte Carlo applications, and in some cases can
undermine the procedure enough to cause convergence to the wrong result.

The Mersenne Twister~\cite{Matsumoto1998a} is the \textsl{de facto} standard
for statistical applications and is used by default in packages such as
MATLAB. It features a large state size and long period, and has recently been
ported to GPUs~\cite{Saito2011}. However, it has a fixed and perhaps 
over-large state size, and
is difficult to tune for optimal performance on GPUs. 
In this work we adapt the
xorgens algorithm of \cite{Brent2007,Brent2008}. 
The attraction of this
approach is the flexible choice of period and state size, facilitating
the optimisation of speed and statistical quality within the resource
constraints of a particular GPU architecture.

We begin with a brief overview of CUDA, then discuss qualitative testing of
PRNGs, including the Mersenne Twister for Graphic Processors (MTGP), CURAND
and xorgens generators.  We then describe our adaptation of the xorgens
algorithm for GPUs. Finally, the results of testing these generators are
presented and some conclusions drawn.

\subsection{The NVIDIA Compute Unified Device Architecture (CUDA) and the Graphics Processing Unit (GPU)}
\label{ssec:cuda}

The Compute Unified Device Architecture (CUDA) was introduced by the NVIDIA
Corporation in November 2006 \cite{Nvidia2010}.  This architecture provides a
complete solution for general purpose GPU programming (GPGPU), including new
hardware, instruction sets, and programming models. The CUDA API allows
communication between the CPU and GPU, allowing the user to control the
execution of code on the GPU to the same degree as on the CPU.

A GPU resides on a \emph{device}, which usually consists of many \emph{multiprocessors} (MPs), each containing some \emph{processors}. 
Each CUDA compatible GPU device has a globally accessible memory address space that is physically separate from the MPs.
The MPs have a local shared memory space for each of the processors associated with the MP. 
Finally, each processor has its own set of registers and processing units for performing computations.

There are three abstractions central to the CUDA software programming model, provided by the API as simple language extensions:
\begin{itemize}
\item A hierarchy of \emph{thread} groupings -- a thread being the smallest unit of processing that can be scheduled by the device.
\item Shared memory -- fast sections of memory common to the threads of a group. 
\item Barrier synchronisation -- a means of synchronising thread operations by halting threads within a group until all threads have met the barrier.  
\end{itemize}

Threads are organised into small groups of 32 called \emph{warps} for
execution on the processors, which are Single-Instruction Multiple-Data (SIMD)
and implicitly synchronous.  These are organised for scheduling across the MPs
in \emph{blocks}.  Thus, each block of threads has access to the same shared
memory space.  Finally, each block is part of a grid of blocks that represents
all the threads launched to solve a problem.  These are specified at the
invocation of \emph{kernels} -- functions executed on the GPU device --
which are managed by ordinary CPU programs, known as \emph{host} code.

As a consequence of the number of in-flight threads supported by a device, and
the memory requirements of each thread, not all of a given GPU device's
computational capacity can be used at once.  The fraction of a device's
capacity that can be used by a given kernel is known as its \emph{occupancy}.

\subsection{Statistical Testing: TestU01}
\label{sec:quality}
Theoretically, the performance of some PRNGs on certain statistical tests
can be predicted, but usually this only applies if the test is performed
over a complete period of the PRNG.  In practice, statistical testing of
{PRNG}s over realistic subsets of their periods requires
empirical methods~\cite{LEcuyer2007,Marsaglia1996}.

For a given statistical test and PRNG to be tested, a test statistic is
computed using a finite number of outputs from the PRNG.  It is required that
the distribution of the test statistic for a sequence of uniform,
independently distributed random numbers is known, or at least that a
sufficiently good approximation is computable~\cite{Leopardi2009}.  Typically,
a {\em $p$-value} is computed, which gives the probability that the test
statistic exceeds the observed value.

The $p$-value can be thought of as the probability that the test statistic
or a larger value would be observed for perfectly uniform and independent
input. Thus the $p$-value itself should be distributed uniformly on $(0,1)$.
If the $p$-value is extremely small, for example of the
order $10^{-10}$, then the {PRNG} definitely {\em fails} the test.
Similarly if $1-p$ is extremely small.
If the $p$-value is not close to $0$ or $1$, then 
the PRNG is said to {\em pass} the test, although this only says that
the test failed to detect any problem with the {PRNG}.

Typically, a whole battery of tests is applied, so that there are many
$p$-values, not just one.  We need to be cautious in interpreting the results
of many such tests; if performing $N$ tests, it is not exceptional
to observe that a $p$-value is smaller than $1/N$ or larger than $1-1/N$. The
TestU01 library presented by \citet{LEcuyer2007} provides a thorough suite of
tests to evaluate the statistical quality of the sequence produced by a
PRNG. It includes and improves on all of the tests in the earlier DIEHARD
package of Marsaglia~\cite{Marsaglia1996}.

\subsection{The Mersenne Twister for Graphic Processors}
The MTGP generator is a recently-released variant of the well known Mersenne
Twister~\cite{Matsumoto1998a,Saito2011}.  As its name suggests, it was
designed for GPGPU applications.  In particular, it was designed with parallel
Monte Carlo simulations in mind.  It is released with a parameter generator
for the Mersenne Twister algorithm to supply users with distinct generators on
request (MTGPs with different sequences).  The MTGP is implemented in NVIDIA
CUDA~\cite{Nvidia2010} in both 32-bit and 64-bit versions.  Following the
popularity of the original Mersenne Twister PRNG, this generator is a suitable
standard against which to compare GPU-based PRNGs.

The approach taken by the MTGP to make the Mersenne Twister parallel can be
explained as follows.
The next element of the sequence, ${x}_i$, is
expressed as some function, $h$, of a number of previous elements in the
sequence, say
\begin{equation*}
{x}_i = h({x}_{i-N},{x}_{i-N+1},{x}_{i-N+M}).  
\end{equation*} 
The parallelism that can be exploited in this algorithm becomes apparent
when we consider the pattern of dependency between further elements of the
sequence:
\begin{eqnarray*}
{x}_i 					&=\;& h({x}_{i-N},{x}_{i-N+1},{x}_{i-N+M}) 		\\ 
{x}_{i+1} 			&=\;& h({x}_{i-N+1},{x}_{i-N+2},{x}_{i-N+M+1})	\\
												& \vdots &						 		 																					\\
{x}_{i+N-M-1} 	&=\;& h({x}_{i-M-1},{x}_{i-M},{x}_{i-1})				\\
{x}_{i+N-M}			&=\;& h({x}_{i-M},{x}_{i-M+1},{x}_{i}).					  
\end{eqnarray*}
The last element in the sequence, which produces ${x}_{i+N-M}$, requires the
value of ${x}_i$, which has not yet been calculated. Thus, only $N-M$ elements
of the sequence produced by a Mersenne Twister can be computed in parallel.

As $N$ is fixed by the Mersenne Prime chosen for the algorithm, all that can
be done to maximise the parallel efficiency of the MTGP is careful selection
of the constant $M$. This constant, specific to each generator, determines the
selection of one of the previous elements in the sequence in the recurrence
that defines the MTGP.  Thus, it has a direct impact on the quality of the
random numbers generated, and the distribution of the sequence.

\subsection{CURAND}
The CUDA CURAND Library is NVIDIA's parallel PRNG framework and library. It
is documented in~\cite{Nvidia2010}. The default generator for this library is
based on the XORWOW algorithm introduced by \citet{Marsaglia2003}.
The XORWOW algorithm is an example of the \emph{xorshift} class of generators.  

Generators of this class have a number of advantages.
The algorithm behind them is particularly simple when compared to other generators such as the Mersenne Twister.
This results in simple generators which are very fast but 
still perform well in statistical tests of randomness.

The idea of the xorshift class generators is to combine two terms in the
pseudo-random sequence (integers represented in binary) using left/right
shifts and ``exclusive or'' (xor) operations to produce the next term in the
sequence. 
Shifts and xor operations
can be performed quickly on computing architectures,
typically faster than operations such as multiplication and division. Also,
generators designed on this principle generally do not require a large
number of values in the sequence to be retained (i.e. a large state space)
in order to produce a sequence of satisfactory statistical quality.

\subsection{xorgens}

Marsaglia's original paper~\cite{Marsaglia2003} only gave xorshift generators
with periods up to $2^{192}-1$. \citet{Brent2007} recently proposed the
\emph{xorgens} family of PRNGs that generalise the idea and have
period $2^n-1$, where $n$ can be chosen to be any convenient power of two up to
$4096$.  The xorgens generator has been released as a free software package,
in a C language implementation (most recently xorgens version
3.05~\cite{Brent2008}).

Compared to previous xorshift generators, the xorgens family has several
advantages:
\begin{itemize}
\item A family of generators with different periods and corresponding
memory requirements, instead of just one.
\item Parameters are chosen optimally, subject to certain criteria designed
to give the best quality output.
\item The defect of linearity over GF(2) is overcome efficiently by
combining the output with that of a Weyl generator.
\item Attention has been paid to the initialisation code (see comments 
in \cite{Brent2007,Brent2008} on proper initialisation), so the generators
are suitable for use in a parallel environment. 
\end{itemize}

For details of the design and implementation of the xorgens family, we
refer to~\cite{Brent2007,Brent2008}. Here we just comment on the combination
with a Weyl generator.

This step is performed to avoid the problem of linearity over GF(2) that
is common to all generators of the Linear-Feedback Shift Register class (such as the Mersenne Twister and CURAND).
A Weyl generator
has the following simple form:
\begin{equation*}
	w_k=w_{k-1} + \omega \mod 2^w,
\end{equation*}
where $\omega$ is some odd constant (a recommended choice is an odd
integer close to $2^{w-1}(\sqrt{5}-1)$). The final output
of an xorgens generator is given by:
\begin{equation} \label{eqn:xor_final}
	w_k(I + R^\gamma) + x_k \mod 2^w,
\end{equation}
where $x_k$ is the output before addition of the Weyl generator,
$\gamma$ is some integer constant close to ${w}/{2}$,
and $R$ is the right-shift operator.  The inclusion of the term $R^\gamma$
ensures that the least-significant bits have high linear complexity (if we
omitted this term, the Weyl generator would do little to improve the quality
of the least-significant bit, since $(w_k \bmod 2)$ is periodic with
period~$2$).

As addition mod $2^w$ is a non-linear operation over GF(2), the result is a
mixture of operations from two different algebraic structures, allowing the
sequence produced by this generator to pass all of the empirical tests
in BigCrush, including those failed by the Mersenne Twister. 
A bonus is that the period is increased by a factor $2^w$ (though this is
not free, since the state size is increased by $w$ bits).

\section{xorgensGP}   


Extending the xorgens PRNG to the GPGPU domain is a nontrivial endeavour, with
a number of design considerations required. We are essentially seeking to
exploit some level of parallelism inherent in the flow of data.  To realise
this, we examine the recursion relation describing the xorgens algorithm:
\begin{equation*} 
	{x}_i = {x}_{i-r}(I + L^a)(I + R^b) + {x}_{i-s}(I + L^c)(I + R^d).
\end{equation*}  
In this equation, the parameter $r$ represents the degree of recurrence, and
consequently the size of the state space (in words, and not counting
a small constant for the Weyl generator and a circular array index). 
$L$ and $R$ represent left-shift and right-shift operators, respectively.
If we
conceptualise this state space array as a circular buffer of $r$ elements we
can reveal some structure in the flow of data. In a circular buffer,
${x}$, of $r$ elements, where $\mathtt{x[i]}$ denotes the
$i^{\text{th}}$ element, ${x}_i$, the indices $i$ and $i+r$
would access the same position within the circular buffer.  This means that
as each new element ${x}_i$ in the sequence is calculated from
$\mathtt{x[i-r]}$ and $\mathtt{x[i-s]}$, the result replaces the
$r^{\text{th}}$ oldest element in the state space, which is no longer
necessary for calculating future elements.

Now we can begin to consider the parallel computation of a sub-sequence of
xorgens. Let us examine the dependencies of the data flow within the buffer
$\mathtt{x}$ as a sequence is being produced:
\begin{eqnarray*}
{x}_i					&=\;& {x}_{i-r}A + {x}_{i-s}B 							\\
{x}_{i+1}			&=\;& {x}_{i-r+1}A + {x}_{i-s+1}B 	
\\  & \vdots & \\ 
{x}_{i+(r-s)}	&=\;& {x}_{i-r+(r-s)}A + {x}_{i-s+(r-s)}B 		\\
 &=\;& {x}_{i-s}A + {x}_{i+r-2s}B 
\\  & \vdots &																								\\
{x}_{i+s}			&=\;& {x}_{i-r+s}A + {x}_{i-s+s}B						\\
											&=\;& {x}_{i-r+s}A + {x}_{i}B.
\end{eqnarray*}
If we consider the concurrent computation of the sequence, we observe that
the maximum number of terms that can be computed in parallel
is \[\min(s, r-s).\]  Here $r$ is fixed by the period required, but
we have some freedom in the choice of $s$.  It is best to choose
$s \approx r/2$ to maximise the inherent parallism.  However, the
constraint ${\rm GCD}(r,s) = 1$ implies that the best we can do is
$s = r/2 \pm 1$, except in the case $r=2$, $s=1$.
This
provides one additional constraint, in the context of xorgensGP versus
(serial) xorgens, on the
parameter set $\{r,s,a,b,c,d\}$ defining a generator.
Thus, we find the thread-level parallelism inherent to the xorgens class of generators.

In the CUDA implementation of this generator we considered the approach of
producing independent subsequences.  With this approach the problem of
creating one sequence of random numbers of arbitrary length, $L$, is made
parallel by $p$ processes by independently producing $p$ subsequences of
length $L/p$, and gathering the results.  With the block of threads
architecture of the CUDA interface and this technique, it is a logical and
natural decision to allocate each subsequence to a block within the grid of
blocks.  This can be achieved by providing each block with its own local
copy of a state space via the shared memory of an MP, and then using the
thread-level parallelism for the threads within this block.  Thus, the local
state space will represent the same generator, but at different
points within its period (which is sufficiently long that overlapping 
sequences are extremely improbable).

Each generator is identical in that only one parameter set $\{r,s,a,b,c,d\}$
is used.  An advantage of this is that the parameters are known at compile
time, allowing the compiler to make optimisations that would not be available
if the parameters were dynamically allocated, and thus known only at runtime.
This results in fewer registers being required by each thread.  For the
generator whose test results are given in \S\ref{sec:results}, we used the
parameters $(r,s,a,b,c,d)=(128,65,15,14,12,17)$. 

\section{Results}\label{sec:results}
We now present an evaluation on the results obtained in our comparison of the
existing GPU PRNG against our implementation of xorgensGP.  All experiments
were performed on a NVIDIA GeForce GTX 480 and a single GPU on the NVIDIA
GeForce GTX 295 (which is a dual GPU device), using the CUDA 3.2 toolkit and
drivers.  Performance results are presented in Table \ref{tab:rng_stat},
and qualitative results in Table \ref{tab:testu01}.
 
\begin{table}
\begin{center}
	\caption[RNG stats]{Approximate memory footprints, 
periods and speed on two devices for 32-bit generators.} 
\begin{tabular}{l r l r r}
	\hline
	\noalign{\smallskip}
	Generator	& State-Space		& Period						& GTX 480 RN/s				&	GTX 295 RN/s 				\tabularnewline 
	\noalign{\smallskip}
	\hline
	\noalign{\smallskip}
	xorgensGP	& $129$ words		&	$2^{4128}$				& $17.7 \times 10^9$	& $ 9.1 \times 10^9$ 	\tabularnewline 
	MTGP 			& $1024$ words	&	$2^{11213}$				& $17.5 \times 10^9$ 	& $10.7 \times 10^9$ 	\tabularnewline 
	CURAND 		& $6$ words     & $2^{192}$ 	& $18.5 \times 10^9$ 	& $ 7.1 \times 10^9$ 	\tabularnewline
	\noalign{\smallskip}
	\hline
\end{tabular}
	\label{tab:rng_stat}
\end{center}
\end{table} 

We first compared the memory footprint of each generator. This depends on the
algorithm defining the generator. The CURAND generator was determined to
have the smallest memory requirements of the three generators compared, and
the MTGP was found to have the greatest.  The MTGP has the longest period
($2^{11213}-1)$, and the CURAND generator has the shortest period
($2^{192}-2^{32}$).

Next, we compared the random number throughput (RN/s) of each generator on
the two different devices.  This was obtained by repeatedly generating 
$10^{8}$ random numbers and timing the duration to produce the sequence of
that length.  We found that the performance of each generator was roughly 
the same, with no significant speed advantage for any generator.  
On the newer GTX 480,
the CURAND generator was the fastest, and the MTGP was the slowest.  
On the older architecture of the
GTX 295 the ordering was reversed: the CURAND generator was the slowest and the
MTGP was fastest. 
These results can be explained in part by the fact that the
CURAND generator was designed with the current generation of ``Fermi'' cards
like the GTX 480, and the MTGP was designed and tested initially on a card
very similar to the GTX 295. In any event, the speed differences are small
and implementation/platform-dependent.

Finally, to compare the quality of the sequences produced, each of the
generators was subjected to the SmallCrush, Crush, and BigCrush batteries of
tests from the TestU01 Library. The xorgensGP generator did not fail any of
the tests in any of the benchmarks.  Only the MTGP failed in the Crush
benchmark, where it failed two separate tests.  This was expected as the
generator is based on the Mersenne Twister, and the tests are designed to
expose the problem of linearity over GF(2).  The MTGP
failed the corresponding, more rigorous tests in BigCrush. Interestingly,
the CURAND generator failed one of these two tests in BigCrush. 

\begin{table}
\begin{center}
\caption[TestU01 Results]{Tests failed in each standard TestU01 benchmark.}
	\label{tab:testu01}
\begin{tabular}{l l l l}
	\hline
	\noalign{\smallskip}
	Generator	& SmallCrush 	& Crush			& BigCrush	\tabularnewline 
	\noalign{\smallskip}
	\hline
	\noalign{\smallskip}
	xorgensGP	& None 				&	None 			& None			\tabularnewline 
	MTGP 			& None				&	\#71,\#72	&	\#80,\#81	\tabularnewline 
	CURAND 		& None				& None   		& \#81 		  \tabularnewline
	\noalign{\smallskip}
	\hline
\end{tabular}
\end{center}
\end{table} 

\section{Discussion}
We briefly discuss the results of the statistical tests,
along with some design considerations for the xorgensGP generator.

CURAND fails one of the
TestU01 tests.  This test checks for linearity and exposes this flaw in the
Mersenne Twister. However, like the xorgensGP, CURAND combines the output of
an xorshift generator with a Weyl generator to avoid linearity over
GF(2), so it was expected to pass the test. 
The period $2^{192}-2^{32}$ of the CURAND generator is much smaller
than that of the other two generators. The BigCrush test consumes 
approximately $2^{38}$
random numbers, which is still only a small fraction of the period.

A more probable explanation relates to the initialisation of the generators
at the block level.  In xorgensGP each block is provided with consecutive
seed values (the id number of the block within the grid).  Correlation
between the resulting subsequences is avoided by the method xorgens uses to
initialise the state space.  It is unclear what steps CURAND takes in its
initialisation.

The MTGP avoids this problem by providing each generator with different
parameter sets for values such as the shift amounts. In developing xorgensGP
this approach was also explored. However, it was found that the overhead of
managing the parameters increased the memory footprint of each generator and
consequently reduced the occupancy and performance of the generator, without
any noticeable improvement on the quality and so was not developed any
further.
 
In conclusion, we presented a new PRNG xorgensGP for GPUs using CUDA. We
showed that it performs with comparable speed to existing solutions and with
better statistical qualities. The proposed generator has a period that
is sufficiently large for statistical purposes 
while not requiring too much state space,
allowing it to give good performance on different devices.
 
\bibliographystyle{abbrvnat}
\bibliography{rpb241}









\end{document}